\documentclass[aps,prl,reprint,groupedaddress]{revtex4-1}

\usepackage{graphicx}
\usepackage{dcolumn}
\usepackage{bm}
\usepackage{hyperref}%
\usepackage{amsmath}
\usepackage{amssymb}

\begin{document}


\title{Generalized individual-based epidemic model for vulnerability assessment of correlated scale-free complex networks}


\author{Mina Youssef}
\author{Caterina Scoglio}
\affiliation{K-State Epicenter, Department of Electrical and Computer Engineering, Kansas State University, Manhattan KS 66506}



\begin{abstract}
Many complex networks exhibit vulnerability to spreading of epidemics, and such vulnerability relates to the viral strain as well as to the network characteristics. For instance, the structure of the network plays an important role in spreading of epidemics. Additionally, properties of previous epidemic models require prior knowledge of the complex network structure, which means the models are limited to only well-known network structures. In this paper, we propose a new epidemiological SIR model based on the continuous time Markov chain, which is generalized to any type of network. The new model is capable of evaluating the states of every individual in the network. Through mathematical analysis, we prove an epidemic threshold exists below which an epidemic does not propagate in the network. We also show that the new epidemic threshold is inversely proportional to the spectral radius of the network. In particular, we employ the new epidemic model as a novel measure to assess the vulnerability of networks to the spread of epidemics. The new measure considers all possible effective infection rates that an epidemic might possess. Next, we apply the measure to correlated networks to evaluate the vulnerability of disassortative and assortative scale-free networks. Ultimately, we verify the accuracy of the theoretical epidemic threshold through extensive numerical simulations. Within the set of tested networks, the numerical results show that disassortative scale-free networks are more vulnerable to spreading of epidemics than assortative scale-free networks.
\end{abstract}


\maketitle

\section{I. Introduction}

Complex networks, such as social networks \cite{WS:98,ASBS:00,LEASA:01,GN:02}, food-webs \cite{S:01}, biological networks \cite{JTAOB:00,JMBO:01,FW:00,WM:00,MS:02}, the world-wide-web (WWW), and the Internet, largely represent many systems from a topological structure point of view. Inherently, a complex network has many dynamics that describe the state of the system and its functionality. Among these dynamics, the spread of epidemics process has attracted the attention of many researchers in multidisciplinary fields. Essentially, epidemics, like human contagious, are represented by the standard compartmental epidemiological model so-called susceptible/infected/removed (SIR), which emulates the spread process. 

The SIR model analytically reveals how an individual'state is changed among the three SIR states in the complex networks. To clarify, during the spread of an epidemic, an individual is in one of the three SIR states. First, a susceptible individual can receive the infection from an infectious neighbor and become infected. Accordingly, the infected individual becomes infectious with infection rate $\beta$. Also, an infected individual can cure itself with a cure rate $\delta$. The curing process represents either the death (removal) or the complete recovery of the individual after the infection. Additionally, the ratio between $\beta$ and $\delta$ is called the effective infection rate. An epidemic threshold $\tau$ is a specific value of the effective infection rate above which an epidemic outbreak takes place. Moreover, it is a function of the network characteristics. 
\newline
Different SIR models are applied to some classes of complex networks ~\cite{BPV:03,N:02a,MSV:02,BBSV:05,VM:09,FDP:09,KR:07,ML:01,YWRBSWZ:07,BMNP:09,TTV:98} depending on the network characteristics. Early SIR models are homogeneous, i.e., all individuals have a similar probability of being infected and infectious. On the other hand, SIR models are also applied on structured networks considering the local connectivity of the network's individuals. For example, scale-free (SF) networks, which are networks owning power-law node degree distribution $P(k) \sim k^{-2-\nu}$ with $0<\nu\leq1$, show a high level of vulnerability to the spreading of epidemics due to highly heterogeneous node degrees distribution when the minimum node degree is greater than two ~\cite{BPV:03}. In addition, the spread of epidemics was studied on correlated networks and uncorrelated networks separately. Thus, we concluded that properties (e.g. epidemic threshold) of previous models are not generalized, and therefore the properties depend on the network structure (e.g. scale-free, small-world, correlated, uncorrelated, regular, exponential, ... etc). Moreover, given the network topology and a suitable SIR model, assessing the vulnerability of the network with respect to the spread of epidemics is difficult. In fact, the epidemic threshold is not always a complete vulnerability measure since it is a binary indication of the epidemic outbreak, and it does not account for the number of infected individuals. Moreover, it has been proven that an epidemic prevails on any SF network regardless of its node degree correlation due to the absence of the epidemic threshold in the limit of a very large number of individuals residing in the network ~\cite{BPV:03}. However, within the class of SF networks, correlated and uncorrelated topologies exist, and they behave differently with respect to spread of epidemics. Therefore, the epidemic threshold is not an adequate measure, and consequently, vulnerability assessment becomes a tough task. 
\newline
In this paper, we propose a novel network vulnerability assessment method. The new method considers all possible effective infection strengths that can harm the network. We focus our study on SF correlated networks to evaluate the vulnerability of disassortative and assortative SF networks. Next, we present a novel individual-based SIR model, which is inspired by the Markov chain approach. We separately study the state of each individual during the infection process, revealing the role of the individual's local connectivity in spreading the infection across the network. Although the exact SIR model, based on the Markov chain stochastic process, describes the global change in the state probabilities of the network, it is limited to small networks due to the exponential divergence in the number of possible network states $3^{N}$ with the growth of network size $N$. Instead, our new model aims to reduce the complexity of the problem and to offer insights into the epidemic spreading mechanism. Through the new SIR model, we study the spread of epidemics on any type of network regardless of its topological structure. Finally, we analytically derive the epidemic threshold for the new model. We find that the new epidemic threshold is inversely proportional to the spectral radius $\lambda_{max}$ (the supremum eigenvalue within the eigenvalue spectrum) of the network. We perform extensive simulations to validate the new SIR model and the new epidemic threshold. Quantitatively, we show that disassortative SF networks are more vulnerable to the spread of epidemics than are assortative SF networks given the same number of individuals $N$ and the same number of connections $L$.
\newline
The paper is organized as follows. In Sec. II, we shed some light on the homogeneous and heterogeneous mixing hypothesis models that exist in the literature. In Sec. III, we introduce the new individual-based SIR model, and we show how it is inspired by the Markov chain model. We also derive the new epidemic threshold, we determine the condition of the existence of a maximum value of number of infected individuals, and we study the role of the network eigenvalue spectrum on the spread of epidemics. In Sec. IV we introduce the new vulnerability assessment of any complex networks, and validate and discuss our analytical findings through extensive simulations. Finally, in Sec. V we conclude our work.

\section{II. SIR Model}
\label{sec:model}

The science of the spread of epidemics is based on compartmental models that assume individuals are classified into non-intersecting sets ~\cite{AM:92,M:93}. Thus, the classical susceptible/infected/removed SIR model characterizes diseases that lead to either immunization or death of individuals. The infected individuals are in the infected set, the healthy ones are in the susceptible set, and the cured or removed ones are in the removed set. Initially, a small number of infected individuals exist that try to infect their susceptible (healthy) neighbors. After receiving the infection, susceptible individuals become infected, and later they try to infect their susceptible neighbors. In this case, infected individuals are infectious. Subsequently, every infected individual either is cured due to immunization or removed due to death. This process was early described by the homogeneous mixing SIR model ~\cite{AM:92}, which evaluates the change in the susceptible $s(t)$, such that infected $i(t)$ and removed $r(t)$ population densities with time, while preserving the overall density at any time $t$, $s(t)+i(t)+r(t)=1$. In the homogeneous mixing model, the rates of changes in densities are governed by the following continuous time differential equations:

\begin{eqnarray}
\frac{ds(t)}{dt} & = & -<k> \beta i(t) s(t), \\
\frac{di(t)}{dt} & = & -\delta i(t) + <k> \beta i(t) s(t), \\
\label{eq:hrt}
\frac{dr(t)}{dt} & = & \delta i(t).
\end{eqnarray}

These differential equations interpret the infection and cure processes. Initially, the spreading process starts with a small infected density $i(0)\simeq0$, the susceptible density is almost one $s(0)\simeq1$, and the removed density is zero $r(0)=0$. Every infected individual infects on average $<k>$ susceptible neighbors, each with an infection rate $\beta$, where $<k>=\sum_{d}dp(d)$ is the average node degree (average number of contacts), and $p(d)$ is the probability of having an individual with degree $d$. Following the differential Eq. (\ref{eq:hrt}), an infected individual is removed at a rate $\delta$. The removed density increases with time until it reaches a certain density level depending on the strength of the epidemic. A non-zero epidemic threshold exist and it is equal to $<k>^{-1}$. If the effective infection rate $\frac{\beta}{\delta}$ is above the threshold, the epidemic prevails in the network. On the other hand, if the effective infection rate is below the threshold, the infected density is very small in the thermodynamic limit. Since on average every infected individual infects a constant number of neighbors, the homogeneous model does not count the heterogeneity in the node degrees of individuals in the network.  
\newline
Another model in the literature is the heterogeneous mixing SIR model ~\cite{MSV:02,BPV:03}, which was proposed to overcome the shortcomings of the homogeneous model. In this model, individuals are classified according to their node degrees. Thus, for a given node degree $d$, the states' densities  $s_{d}(t)$, $i_{d}(t)$and $r_{d}(t)$ evolve with time $t$, and their sum is constant, such that $s_{d}(t)+i_{d}(t)+r_{d}(t)=1$. The rates of changes in the three states for a given node degree $d$ are governed by the following set of differential equations:

\begin{eqnarray}
\frac{ds_{d}(t)}{dt} & = & -d \beta s_{d}(t) \theta(t),\\
\frac{di_{d}(t)}{dt} & = & -\delta i_{d}(t) + d \beta s_{d}(t) \theta(t),\\
\frac{dr_{d}(t)}{dt} & = & \delta i_{d}(t).
\end{eqnarray}

The probability that a link is pointing to an infected individual is given by the factor $\theta(t)$, where $\theta(t)$ is found to be $\frac{\sum_{d}d p(d)i_{d}(t)}{<k>}$. 
This model was applied to both uncorrelated and correlated complex networks, leading to further analysis of the epidemic threshold. For uncorrelated networks, the epidemic threshold is $\tau^{ucr}=\frac{<k>}{<k^{2}>-<k>}$, where $<k^{2}>$ is the second moment of the node degrees. On the other hand, the epidemic threshold for correlated networks is $\tau^{cr}=\frac{1}{\overline{\Lambda}_{m}}$, where $\overline{\Lambda}_{m}$ is the maximum eigenvalue of the connectivity matrix $\overline{C}_{dd'}=\frac{d(d'-1)}{d'}p(d'\mid d)$. 
Although this model considers the heterogeneous connectivity in the networks, it does not reveal the state of each individual in the network. It only reflects the evolution of the densities over time for a given node degree, while neglecting the states of individuals within the same node degree.

\section{III. Individual-based SIR model}
\label{sec:individualbased}

In this paper, we present a new individual-based SIR model in which each node can be either susceptible $S$, infected $I$ or recovered $R$ with a given probability for each state. The new model is inspired by the continuous time Markov chain SIR model. However, instead of considering the combinatorial states of the individuals in the network, we study each individual deliberately \cite{MOK:09}, by decomposing the infinitesimal $Q_{3^{N} \times 3^{N}}$ matrix to $N$ infinitesimal matrices, each with three states as follows:

\begin{eqnarray}
q_{k}(t)=
\left[
\begin{array}{ccc}
-\beta \sum_j a_{k,j} 1_{[i_{j}(t)=1]} & \beta \sum_j a_{k,j} 1_{[i_{j}(t)=1]} & 0 \\
0 & -\delta & \delta \\
0 & 0 & 0
\end{array}\right]
\nonumber
\end{eqnarray}

where $a_{k,j}$ is the binary entry in the network adjacency matrix, representing the existence of a contact between individual $k$ and individual $j$, and the indicator function $1_{[i_{j}(t)=1]}=1$ represents the event that individual $j$ is infected and zero otherwise. In this model, we replace the actual event with its effective probability, and therefore the event $i_{j}(t)=1$ is replaced by $I_{j}(t)=p(i_{j}(t)=1)$.
For every individual $k$, we derive the system of differential equations as follows:

\begin{equation}
\frac{dState_{k}(t)}{dt}=q_{k}^{T}(t) State_{k}(t)
\end{equation}

where $q_{k}^{T}(t)$ is the transpose of $q_{k}(t)$. The obtained differential equations are

\begin{eqnarray}
\label{eq:Seq}
\frac{dS_{k}(t)}{dt} & = & -S_{k}(t) \beta \sum_{j} a_{k,j} I_{k}(t), \\
\label{eq:Ieq}
\frac{dI_{k}(t)}{dt} & = & S_{k}(t) \beta \sum_{j} a_{k,j} I_{k}(t) - \delta I_{k}(t), \\
\label{eq:Req}
\frac{dR_{k}(t)}{dt} & = & \delta I_{k}(t).
\end{eqnarray}

\begin{figure}
\centering
\includegraphics[width=9.5cm]{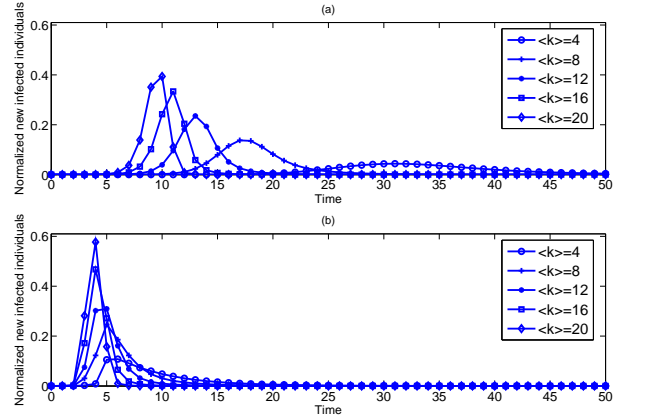}
\caption{\label{fig:timePlot} Normalized new infected individuals as a function of time for $\beta=0.1$ and $\delta=0.2$ given correlated networks with $N=10^{4}$ and different average node degree $<k>$. Two different types of correlated networks are simulated (a) assortative SF networks, and (b) disassortative SF networks. The peak of the new infected individuals in disassortative networks leads to the corresponding peak in asssortative networks.}
\end{figure}

At any time $t$, each individual will be in any of the states with total probability of 1, $S_{k}(t)+I_{k}(t)+R_{k}(t)=1$. In addition, the sum of rates of changes in the state probabilities is zero $\frac{dS_{k}(t)}{dt}+\frac{dI_{k}(t)}{dt}+\frac{dR_{k}(t)}{dt}=0$. Therefore, we only solve $2N$ simultaneous differential equations instead of $3N$. Figure ~\ref{fig:timePlot} shows the time evolution of new infected individuals in assortative and disassortative SF networks with different $<k>$=4, 8, 12, 16 and 20 given $\beta=0.1$ and $\delta=0.2$.  

\subsection{Steady-state population}

To evaluate the behavior of the system of differential equations at the steady state, we equate the differential Eqs. in (\ref{eq:Seq} - \ref{eq:Req}) to zero. The steady-state probability of infection $I_{\infty}$ is always zero, while the steady-state probability of recovery $R_{\infty}$ always have a positive value, which is $\delta \int_{0}^{t_{I_{new}}=0} u^{T}I(z) dz$ where $t_{I_{new}}=0$, the time at which there are no more new infected individuals in the network, and $u^{T}$ is the transpose of a vector of 1's. On the other hand, the steady-state probability of being susceptible $S_{\infty}$ is zero if, and only if, $R_{\infty}=1$, otherwise, it is a positive value.

\subsection{Epidemic threshold}

The epidemic threshold is the condition that the epidemic prevails in the network. To compute the threshold, we follow the analysis presented in \cite{BPV:03}. We assume that the initial fraction of infected individuals is very small and therefore $S_{k}(0) \backsimeq 1$. The differential Eq. (\ref{eq:Ieq}) is written as follows:

\begin{equation}
\frac{dI_{k}(t)}{dt} \backsimeq \sum_{j} \tilde{L_{k,j}} I_{j}(t)
\label{eq:approxdifeq}
\end{equation}
 
where the element $\tilde{L}_{k,j}=\beta a_{k,j} - \delta\delta_{k,j}$ is the entry of the Jacobian matrix $\tilde{L}=\{\tilde{L}_{k,j}\}=\beta A - \delta I_{N \times N}$, and $\delta_{k,j}$ is the Kronecker delta function and equals 1 forall $k=j$. Since any element $a_{k,j}$ of the symmetric adjacency matrix $A$ is either 0 or 1, and according to Frobenius theorem, the maximum eigenvalue $\lambda_{max,A}$ of $A$ is positive and real, the eigenvalues of the matrix $\tilde{L}$ have the form of $\beta \lambda_{i,A} - \delta$, and the eigenvectors are the same as those for the adjacency matrix $A$. Thus, the stability condition of the solution $I=0$ of the differential Eq. (\ref{eq:approxdifeq}) is $-\delta +\beta \lambda_{max,A}<0$, and the SIR threshold for any undirected network becomes:

\begin{equation}
\frac{\beta}{\delta} < \frac{1}{\lambda_{max,A}} = \tau
\label{eq:threshold}
\end{equation}

The threshold states that whenever $\frac{1}{\lambda_{max,A}}$ is greater than the effective infection rate $\frac{\beta}{\delta}$, an epidemic does not prevail in the network.

\subsection{The existence of a maximum number of infected individuals}

The number of infected individuals increases in time following a certain profile ~\cite{BBV:08} depending on the infection strain. Below, we derive the condition for which a maximum number of infected individuals occurs, and how the condition is related to the epidemic threshold.
Let $u^{T}I(t)=\sum_{k}I_{k}(t)$ be the total number of infected individuals in the network. The existence of a maximum value for $I(t)$ is determined through $\frac{du^{T}I(t)}{dt}=\sum_{k}\frac{dI_{k}(t)}{dt} = 0$, and we obtain:

\begin{equation}
\label{eq:maxI}
\sum_{k} \left[ S_{k}(t) \beta \sum_{j} a_{k,j} I_{j}(t) - \delta I_{k}(t)\right] = 0
\end{equation}

By rewriting Eq. (\ref{eq:maxI}) in the matrix form, we obtain the following equation:

\begin{equation}
\label{eq:maxzero}
\left[\beta S^{T}(t) A - \delta u^{T} \right]I(t) = 0
\end{equation}

Eq. (\ref{eq:maxzero}) suggests the possible solutions for $I(t)$ are either $I(t)$ equals zero, which happens at the steady state, or $\beta S^{T}(t) A - \delta u^{T}$ equals zero. The second solution derives a condition for the existence of a positive maximum value of $I(t)$. Consequently, the second solution $A S(t) = \frac{\delta}{\beta} u$ is on the form of $Wx=\rho x$, where $x$ and $\rho$ are an eigenvector and an eigenvalue of the matrix $W$, respectively. The vector $S(t)$ is equal to the vector $u$ only if $\frac{\delta}{\beta}$ is equal to the maximum eigenvalue $\lambda_{max,A}$ of $A$, which follows Frobenius theorem and takes place for $t \to 0$ and $S(0) \to 1$. Moreover, this solution proves the existence of the epidemic threshold shown in inequality (\ref{eq:threshold}) whenever $\frac{\delta}{\beta} < \lambda_{max,A}$, and therefore the epidemic spreads in the network, and $S_{k}(t)\leq 1$ forall $k$.

\subsection{The effect of the network spectrum}

To address the effect of the spectrum of the adjacency matrix $A$, we write the rate of change as a total fraction of infected individuals $u^{T}I(t)$ as follows: 

\begin{equation}
\frac{du^{T}I(t)}{dt} = \beta (u^{T}-I^{T}(t)-R^{T}(t)) A I(t) - \delta u^{T}(t) I(t).
\label{eq:spectrum1}
\end{equation} 

Denote the vector of node degrees $D=u^{T} A$, and the eigenvalue decomposition of the adjacency matrix $A=U\Lambda U^{T}$. We rewrite the differential Eq. (\ref{eq:spectrum1}) as follows:

\begin{eqnarray}
\frac{du^{T}I(t)}{dt} = && (\beta D - \delta u)^{T} I(t) - \beta (U^{T}I(t))^{T} \Lambda (U^{T}I(t)) \nonumber \\
&& - \beta (U^{T}R(t))^{T} \Lambda (U^{T}I(t))
\label{eq:spectrum2}
\end{eqnarray}

Let $x_{j}$ be the $j^{th}$ element in the vector $U^{T}I(t)$, and let $y_{j}$ be the $j^{th}$ element in the vector $U^{T}R(t)$. We rewrite the differential equation as follows:

\begin{equation}
\frac{du^{T}I(t)}{dt} = (\beta D - \delta u)^{T} I(t) - \beta \sum_{j=1}^{N} \lambda_{j} x_{j}^{2} - \beta \sum_{j=1}^{N} \lambda_{j} x_{j} y_{j}
\label{eq:spectrum3}
\end{equation}

To relate $I_{max}$ with the spectrum $\lambda_{j}$ and the eigenvectors $U$, let $\frac{du^{T}I(t)}{dt}$ equal zero, and therefore we obtain the following equation:

\begin{equation}
\sum_{k=1}^{N} (d_{k} - \frac{\delta}{\beta}) I_{k_{max}}=\sum_{j=1}^{N} \lambda_{j} x_{j}^{2} - \sum_{j=1}^{N} \lambda_{j} x_{j} y_{j} 
\label{eq:spectrum4}
\end{equation}

Since the matrix $A$ is symmetric, we can see that $\lambda_{max}$ is a positive eigenvalue and therefore the dominant eigenvalue within the spectrum, and elements of the corresponding eigenvector are positive as well. Eq. (\ref{eq:spectrum4}) states that as $\delta$ decreases, the LHS increases, and so $I_{max}$ increases with the eigenvectors corresponding to $\lambda_{max}$, while on the other hand, the corresponding $R$ decreases.

\section{IV. Vulnerability measure}
\label{sec:vulnerability}

\begin{figure}
\includegraphics[width=9.0cm]{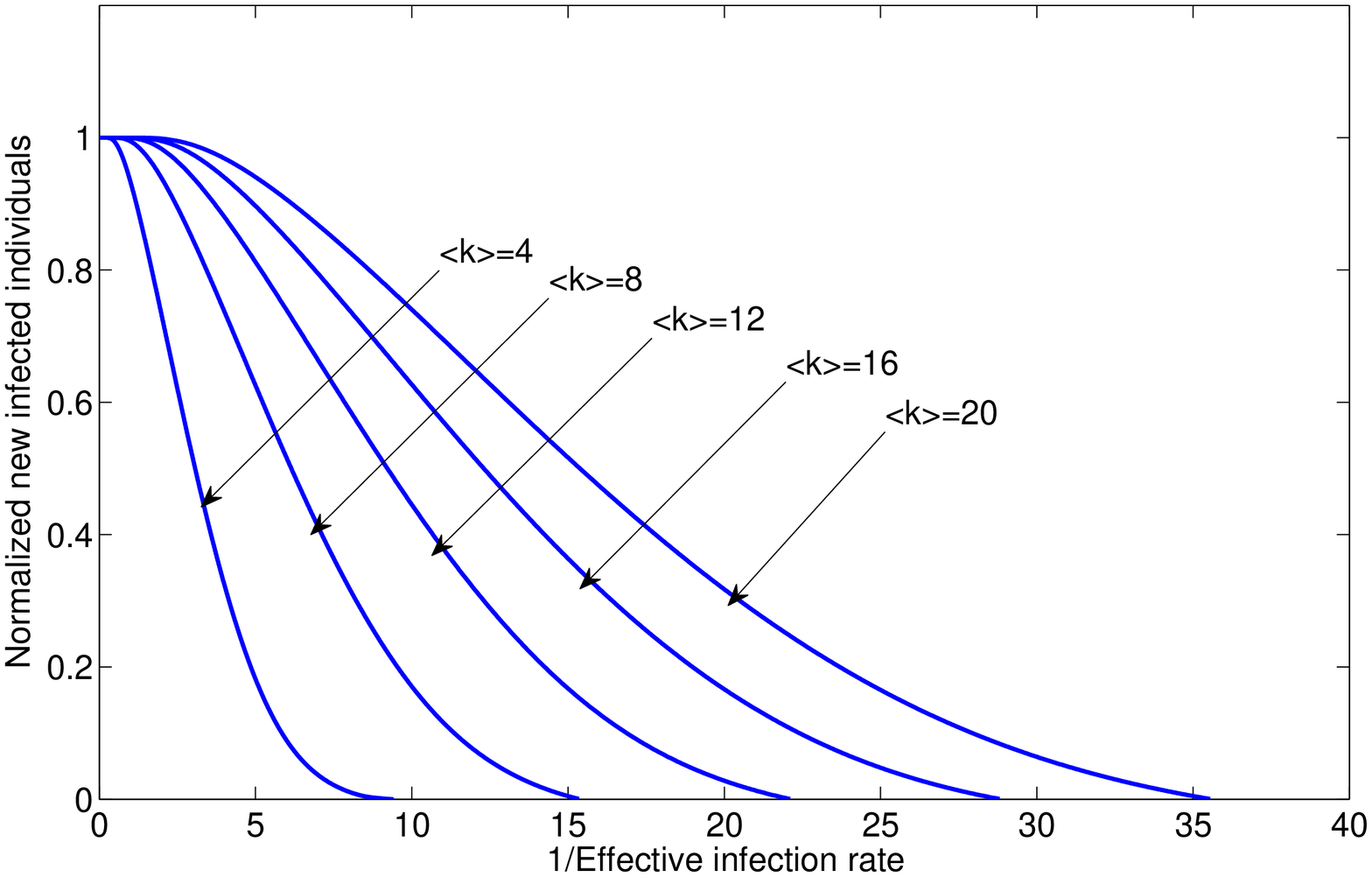}
\caption{\label{fig:assortEpsy} Normalized total of new infected individuals as a function of the inverse of effective infection rate for assortative SF networks given $N=10^{4}$ and different average node degree $<k>$. The curve starts from the point where $\frac{\delta}{\beta}=0$, and the normalized total new infected cases is 1, and then it decreases until it reaches the value zero when the value of $\frac{\delta}{\beta}$ equals the spectral radius of the network.}
\end{figure}

\begin{figure}
\includegraphics[width=9.0cm]{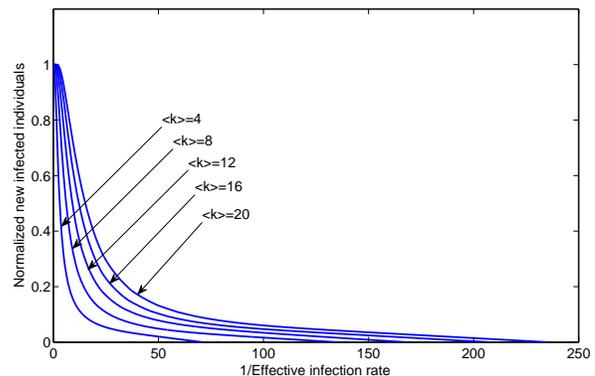}
\caption{\label{fig:disassortEpsy} Normalized total of new infected individuals as a function of the inverse of effective infection rate for disassortative SF networks given $N=10^{4}$ and different average node degree $<k>$. The curve starts from the point where $\frac{\delta}{\beta}=0$, and the normalized total new infected cases is 1, and then it decreases until it reaches the value zero when the value of $\frac{\delta}{\beta}$ equals the spectral radius of the network.}
\end{figure}

We employ the individual-based SIR model to assess the vulnerability of a complex network such that the total number of new infected individuals reflects the vulnerability of the network to the spread of epidemics given any infection strength. In this section, we introduce a new vulnerability assessment measure $\Psi$ with respect to the spread of epidemics, and we define it as the ability of an epidemic to prevail in a complex network given all possible effective infection rates.
Mathematically, we define the assessment measure $\Psi$ by fixing $\beta=\frac{1}{\lambda_{max,A}}$ and for a given cure rate $\delta$, the total number of new infected individuals is  $\int_{0}^{t_{I_{new}=0}}\sum_{k} S_{k}(t) \beta \sum_{j} a_{k,j} I_{k}(t,\delta) dt$. By integrating over the defined range of cure rate $0 \leq \delta \leq 1$, we obtain $\Psi$ as follows:

\begin{equation}
\label{eq:psi}
 \Psi = \int_{0}^{1} \int_{0}^{t_{I_{new}=0}}\sum_{k} S_{k}(t) \beta \sum_{j} a_{k,j} I_{k}(t,\delta) dt d\delta
\end{equation} 
Figures ~\ref{fig:assortEpsy} and ~\ref{fig:disassortEpsy} show the numerical simulations of the spread of an epidemic for $0\leq\frac{\delta}{\beta}\leq \lambda_{max,A}$ on assortative and disassortative SF networks given different average node degrees $<k>$, where $\frac{\delta}{\beta}$ is the inverse of the effective infection rate. 

\begin{figure}
\includegraphics[width=9.0cm]{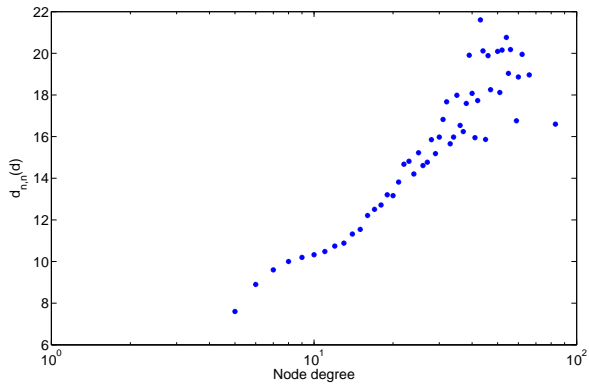}
\caption{\label{fig:assort8knnkPlot} Node degree as a function of average neighbors connectivity $d_{n,n}(d)$ of individuals with the same node degree for a sample of an assortative SF network with $N=10^{4}$ and $<k>=8$. The node degree correlation is an increasing function for an assortative network.}
\end{figure}

\begin{figure}
\includegraphics[width=9.0cm]{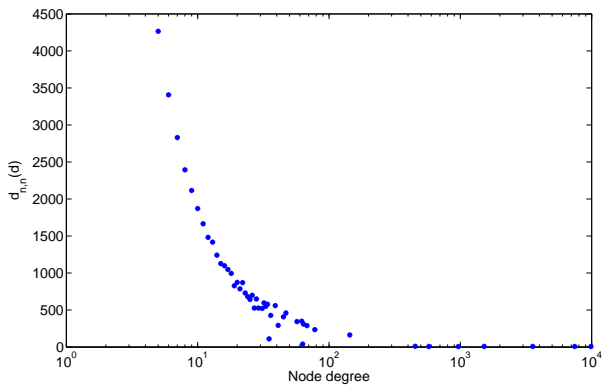}
\caption{\label{fig:disassort8knnkPlot} Node degree as a function of average neighbors connectivity $d_{n,n}(d)$ of individuals with the same node degree for a sample of a disassortative SF network with $N=10^{4}$ and $<k>=8$. The node degree correlation is a decreasing function for a disassortative network.}
\end{figure}

Previous work ~\cite{KSSY:09} introduced a measure that takes into account the number of infected individuals at steady state for the susceptible/infected/susceptible compartmental model. We use the new measure $\Psi$ to evaluate the vulnerability of correlated networks, these in which node degree correlation is observed. They are also classified as assortative and disassortative networks. For example, social networks are classified as assortative networks, while technological and biological networks are classified as disassortative networks ~\cite{N:02b}. In assortative networks, individuals of small node degree are connected with other individuals of small node degree, while individuals with large node degree are connected with other individuals with large node degree. On the other hand, the opposite is true for disassortative networks. 
Pearson assortativity coefficient ~\cite{N:02b,SVV:03} was proposed to characterize the node degree correlation numerically. However, it does not give an accurate measure for networks with complicated degree correlation functions. To accurately describe the degree correlations, we evaluate the average connectivity of the neighbors of an individual $k$ by following the technique presented in ~\cite{SVV:01,VSV:02,BBV:08}:

\begin{equation}
d_{n,n,k}=\frac{1}{d_{k}}\sum_{j\in neighbors(k)} d_{j}
\end{equation}
The average connectivity of neighbors of an individual is averaged overall of all individuals for a given node degree $d$, 
\begin{equation}
d_{n,n}(d)=\frac{1}{N_{d}}\sum_{k/d_{k}=d}d_{n,n,k}
\end{equation}
where $N_{d}$ is the number of individuals of degree $d$. Figures ~\ref{fig:assort8knnkPlot} and ~\ref{fig:disassort8knnkPlot} show two examples for correlated networks, one for an assortative network and the other for a disassortative network, respectively. 

We focus on the vulnerability assessment of correlated SF networks. We generate assortative and disassortative SF networks using the algorithm in ~\cite{GZLBWZ:06}. The algorithm starts with a connected graph with $m_{0} \ll N$ individuals. Every new individual is connected to the already existing individuals through two stages: In the first stage, a new individual is connected to an existing individual $k$ with probability $\pi_{k}=\frac{d_{k}}{\sum_{j}d_{j}}$; in the second stage, a new link between the new individual and one of the neighbors $s$ of the chosen individual $k$ in the first stage is added with probability $p_{s}=\frac{d_{s}^{\alpha}}{\sum_{v\in \Gamma_{k}} d_{v}^{\alpha}}$, where $\alpha$ is an assortative tunning coefficient, and $\Gamma_{k}$ is the set of neighbors of individual $k$ chosen in the first stage.
\newline
To simplify the evaluation of numerical results, both the constructed assortative and disassortative networks have the same number of individuals $N$ and links $L$ with average node degrees $<k>$=4, 8, 12, 16 and 20. Next, we apply the new measure $\Psi$ in Eq. (\ref{eq:psi}) to quantitatively assess the vulnerability of both assortative and disassortative networks.  All the simulations are averaged over 10 runs. Table ~\ref{tab:psimeasure} summarizes the values of $\Psi$ for both types of networks for different average node degrees. We notice that the disassortative networks have higher values of vulnerability measure $\Psi$ than those of assortative networks regardless of the average node degree value. In addition, the $\Psi$ value increases with increases in $L$ (i.e.$<k>$) due to the increase in the effective spreading rate of any infected individual for its susceptible neighbors. 
Moreover, in Fig. ~\ref{fig:timePlot}, we observe that the peaks of normalized new infected individuals in disassortative networks are greater than the peaks in assortative networks; meanwhile, the peaks in disassortative networks lead the corresponding peaks in assortative networks. In other words, an epidemic widely spreads in disassortative networks, and it spread faster than in assortative networks. Fig. ~\ref{fig:timePlot} also reveals insights about any future immunization strategy that could be applied to both networks. For example, we can assume that immunization strategies on assortative and disassortative networks are different. Therefore, in assortative SF networks, mitigtion strategies are going to be more effective than in disassortative SF networks.

\begin{table}
\caption{\label{tab:psimeasure} Vulnerability measure $\Psi$ for assortative and disassortative SF networks given different average node degrees $<k>$. The network size is $N=10^{4}$.}
\begin{ruledtabular}
\begin{tabular}{|c|c|c|}
$<k>$ & Assortative networks & Disassortative networks \cr
\hline
4& 3.32 & 6.54 \cr
8& 6.54 & 12.58 \cr
12& 9.76 & 17.65 \cr
16& 12.98 & 23.47 \cr
20& 16.22 & 28.68 \cr 
\end{tabular}
\end{ruledtabular}
\end{table}

\section{V. Conclusions}
\label{sec:conclusions}

In this paper, we have reviewed the well-known homogeneous and heterogeneous SIR models, and we have shown how both models do not evaluate the state of every individual in the complex networks. To account for this, we have presented a new individual-based SIR model that is derived from the continuous time Markov chain model. The new model evaluates the probability of infection of every individual separately considering the probability of infection of the individual's neighbors. Unlike previous models in the literature whose their properties require a priori knowledge of the topological structure of the network under study, the new individual-based model can be applied to any type of network regardless its structure. We have also derived the epidemic threshold above which an epidemic prevails in the network. We found that the reciprocal of the spectral radius of the complex network is the epidemic threshold showing the role of the network characteristics in the spread of epidemics. In addition, we have shown the condition for the existence of a maximum number of new infected individuals, and how it is related to the epidemic threshold. Moreover, we have shown that the spectral radius and its corresponding eigenvector of the complex network and the effective infection rate determine the maximum number of the new infected individuals. Furthermore, we have presented a new technique $\Psi$ to quantitatively measure the vulnerability of any type of network structures. We have applied the new measure on assortative and disassortative SF networks, and through numerical simulations we have shown that disassortative scale-free networks are more vulnerable than assortative scale-free networks.
The new SIR model and its properties could have implications for many systems that are viewed as complex networks, and the new measure could rank different networks based on their vulnerability with respect to spread of epidemics. 

\begin{acknowledgments}
This work was partially supported by National Agricultural Biosecurity Center NABC at Kansas State University. 
\end{acknowledgments}


\begin{thebibliography}{}%
\makeatletter
\providecommand \@ifxundefined [1]{%
 \ifx #1\undefined \expandafter \@firstoftwo
 \else \expandafter \@secondoftwo
\fi
}%
\providecommand \@ifnum [1]{%
 \ifnum #1\expandafter \@firstoftwo
 \else \expandafter \@secondoftwo
\fi
}%
\providecommand \enquote [1]{``#1''}%
\providecommand \bibnamefont  [1]{#1}%
\providecommand \bibfnamefont [1]{#1}%
\providecommand \citenamefont [1]{#1}%
\providecommand\href[0]{\@sanitize\@href}%
\providecommand\@href[1]{\endgroup\@@startlink{#1}\endgroup\@@href}%
\providecommand\@@href[1]{#1\@@endlink}%
\providecommand \@sanitize [0]{\begingroup\catcode`\&12\catcode`\#12\relax}%
\@ifxundefined \pdfoutput {\@firstoftwo}{%
 \@ifnum{\z@=\pdfoutput}{\@firstoftwo}{\@secondoftwo}%
}{%
 \providecommand\@@startlink[1]{\leavevmode\special{html:<a href="#1">}}%
 \providecommand\@@endlink[0]{\special{html:</a>}}%
}{%
 \providecommand\@@startlink[1]{%
  \leavevmode
  \pdfstartlink
   attr{/Border[0 0 1 ]/H/I/C[0 1 1]}%
   user{/Subtype/Link/A<</Type/Action/S/URI/URI(#1)>>}%
  \relax
 }%
 \providecommand\@@endlink[0]{\pdfendlink}%
}%
\providecommand \url  [0]{\begingroup\@sanitize \@url }%
\providecommand \@url [1]{\endgroup\@href {#1}{\urlprefix}}%
\providecommand \urlprefix [0]{URL }%
\providecommand \Eprint[0]{\href }%
\@ifxundefined \urlstyle {%
  \providecommand \doi [1]{doi:\discretionary{}{}{}#1}%
}{%
  \providecommand \doi [0]{doi:\discretionary{}{}{}\begingroup
  \urlstyle{rm}\Url }%
}%
\providecommand \doibase [0]{http://dx.doi.org/}%
\providecommand \Doi[1]{\href{\doibase#1}}%
\providecommand \bibAnnote [3]{%
  \BibitemShut{#1}%
  \begin{quotation}\noindent
    \textsc{Key:}\ #2\\\textsc{Annotation:}\ #3%
  \end{quotation}%
}%
\providecommand \bibAnnoteFile [2]{%
  \IfFileExists{#2}{\bibAnnote {#1} {#2} {\input{#2}}}{}%
}%
\providecommand \typeout [0]{\immediate \write \m@ne }%
\providecommand \selectlanguage [0]{\@gobble}%
\providecommand \bibinfo [0]{\@secondoftwo}%
\providecommand \bibfield [0]{\@secondoftwo}%
\providecommand \translation [1]{[#1]}%
\providecommand \BibitemOpen[0]{}%
\providecommand \bibitemStop [0]{}%
\providecommand \bibitemNoStop [0]{.\EOS\space}%
\providecommand \EOS [0]{\spacefactor3000\relax}%
\providecommand \BibitemShut [1]{\csname bibitem#1\endcsname}%
\end{thebibliography}%


\begin{thebibliography}{1}

\bibitem{Abbott:TB10-1-59}
Peter Abbott.
\newblock {{{UK\TeX} and the Aston archive}}.
\newblock {\em TUGboat}, 10(1):59--60, April 1989.

\bibitem{Golding:1994}
Richard~A. Golding, Darrell D.~E. Long, and John Wilkes.
\newblock The {\emph{refdbms}} distributed bibliographic database system.
\newblock In {\em Proceedings of the Winter Usenix Conference}, San Francisco,
  CA, January 1994.

\bibitem{Greenwade:TB14-3-342}
George~D. Greenwade.
\newblock {{The Comprehensive {\TeX} Archive Network ({\CTAN})}}.
\newblock {\em TUGboat}, 14(3):342--351, October 1993.

\bibitem{Walsh:TB15-3-339}
Norm Walsh.
\newblock {{A World Wide Web interface to {\CTAN}}}.
\newblock {\em TUGboat}, 15(3):339--343, September 1994.

\end{thebibliography}


\begin{thebibliography}{1}


\bibitem{WS:98}
D. J. Watts and S. H. Strogatz,
\newblock {{Collective dynamics of 'small-world' networks}},
\newblock {\em Nature},393, 440--42, (1998).


\bibitem{ASBS:00}
L. A. Amaral, A. Scala, M. Barthelemy and H. E. Stanley,
\newblock {{Classes of small-world networks}},
\newblock {\em Proc Natl Acad Sci U S A}, 97, 11149--11152, (2000).


\bibitem{LEASA:01}
F. Liljeros, C. R. Edling, L. A. N. Amaral, H. E. Stanley and Y.\o Aberg,
\newblock {{The web of human sexual contacts}},
\newblock {\em Nature}, 411, 907--908, (2001).


\bibitem{GN:02}
M. Girvan and M. E. J. Newman,
\newblock {{Community structure in social and biological networks}},
\newblock {\em Proc. Natl. Acad. Sci. U S A}, 99, 7821--7826, (2002).


\bibitem{S:01}
S. H. Strogatz,
\newblock {{Exploring complex networks}},
\newblock {\em Nature}, 410, 268-276 (2001).


\bibitem{JTAOB:00}
H. Jeong, B. Tombor, R. Albert, Z. N. Oltvai, and A.-L. Barab\'asi,
\newblock {{The large-scale organization of metabolic networks}},
\newblock {\em Nature}, 407, 651-654, (2000).


\bibitem{JMBO:01}
H. Jeong, S. Mason, A.-L. Barab\'asi, and Z. N. Oltvai,
\newblock {{Lethality and centrality in protein networks}},
\newblock {\em Nature}, 411, 41-42, (2001).


\bibitem{FW:00}
D. A. Fell and A. Wagner,
\newblock {{The small world of metabolism}},
\newblock {\em Nature Biotechnology}, 18, 1121-1122, (2000).


\bibitem{WM:00}
R. J. Williams and N. D. Martinez,
\newblock {{Simple rules yield complex food webs}},
\newblock {\em Nature}, 404, 180-183, (2000).


\bibitem{MS:02}
J. M. Montoya and R. V. Sol\'e,
\newblock {{Small world patterns in food webs}},
\newblock {\em J. Theor. Bio.}, 214, 405-412, (2002).


\bibitem{BPV:03}
M. Bogu\~n\'a, R. Pastor-Satorras and A. Vespignani,
\newblock {{Epidemic spreading in complex networks with degree correlations}},
\newblock {\em Lect. Notes Phys}, 625, 127-147, (2003).


\bibitem{N:02a}
M. E. J. Newman,
\newblock {{Spread of epidemic disease on networks}},
\newblock {\em Phy. Rev. E}, 66, 16128, (2002).


\bibitem{MSV:02}
Y. Moreno, R. Pastor-Satorras and A. Vespignani,
\newblock {{Epidemic outbreaks in complex heterogeneous networks}},
\newblock {\em Eur. Phys. J. B}, 26, 521-529, (2002).


\bibitem{BBSV:05}
M. Barth\'elemy, A. Barrat, R. Pastor-Satorras, A. Vespignani,
\newblock {{Dynamical patterns of epidemic outbreaks in complex heterogeneous networks}},
\newblock {\em J Theor Biol.}, 235, 275-88, (2005).


\bibitem{VM:09}
E. Volz and L. A. Meyers ,
\newblock {{Epidemic thresholds in dynamic contact networks}},
\newblock {\em J. R. Soc. Interface}, 6, 233--241, (2009).

 
\bibitem{FDP:09}
B. Frank, S. David, and T. Pieter,
\newblock {{Threshold behaviour and final outcome of an epidemic on a random network with household structure}},
\newblock {\em Adv. in Appl. Probab.}, 41, 3, 765-796, (2009).


\bibitem{KR:07}
E. Kenah, JM. Robins,
\newblock {{Network-based analysis of stochastic SIR epidemic models with random and proportionate mixing}},
\newblock {\em J Theor Biol.}, 249, 4, 706-22, (2007).


\bibitem{ML:01}
R. M. May and A. L. Lloyd,
\newblock {{Infection dynamics on scale-free networks}},
\newblock {\em Phys. Rev. E}, 64, 066112, (2001).


\bibitem{YWRBSWZ:07}
R. Yang, B.-H. Wang, J. Ren, W.-J. Bai, Z.-W. Shi, W.-X. Wang, T. Zhou,
\newblock {{Epidemic spreading on heterogeneous networks with identical infectivity}},
\newblock {\em Physics Letters A}, 364, 189.193, (2007).

 	
\bibitem{BMNP:09}
A. J. Black, A. J. McKane, A. Nunes, and A. Parisi,
\newblock {{Stochastic fluctuations in the susceptible-infective-recovered model with distributed infectious periods}},
\newblock {\em Phys. Rev. E}, 80, 021922, (2009).


\bibitem{TTV:98}
H. C. Tuckwell, L. Toubiana, and J.-F. Vibert,
\newblock {{Spatial epidemic network models with viral dynamics}},
\newblock {\em Phys. Rev. E}, 57, 2163, (1998).


\bibitem{AM:92}
R. M. Anderson and R. M. May,
\newblock {{Infectious diseases in humans}},
\newblock {\em Oxford University Press}, Oxford, (1992).


\bibitem{M:93}
J. D. Murray,
\newblock {{Mathematical Biology}},
\newblock {\em Springer Verlag}, Berlin, (1993).


\bibitem{MOK:09}
P. Van Mieghem, J. S. Omic and R. E. Kooij,
\newblock {{Virus Spread in Networks}},
\newblock {\em IEEE/ACM Transaction on Networking}, Vol. 17, No. 1, pp. 1-14, February 2009.


\bibitem{BBV:08}
A. Barrat, M. Barth\'elemy and A. Vespignani,
\newblock {{dynamical processes on complex networks}},
\newblock {\em Cambridge University Press}, Cambridge (2008).


\bibitem{KSSY:09}
R. Kooij, P. Schumm, C. Scoglio and M. Youssef,
\newblock {{A new metric for robustness with respect to virus spread}},
\newblock {\em IFIP Networking 2009}, Aachen, Germany (2009).


\bibitem{N:02b}
M. E. J. Newman,
\newblock {{Assortative Mixing in Networks}},
\newblock {\em Phys. Rev. Lett.}, 89, 208701, (2002).


\bibitem{SVV:03}
S.N. Dorogovtsev,
\newblock {{Networks with given correlations}},
\newblock {\em arXiv:cond-mat/0308336v1}, (2003).


\bibitem{SVV:01}
R. Pastor-Satorras, A. V\'azquez and A. Vespignani,
\newblock {{dynamical and correlation properties of the Internet}},
\newblock {\em Phys. Rev. Lett.}, 87, 258701, (2001).


\bibitem{VSV:02}
A. V\'azquez, R. Pastor-Satorras and A. Vespignani,
\newblock {{Large-scale topological and dynamical properties of the Internet}},
\newblock {\em Phys. Rev. E}, 65, 066130, (2002).


\bibitem{GZLBWZ:06}
Q. Guo, T. Zhou, J.-G. Liu, W.-J. Bai, B.-H. Wang and M. Zhao,
\newblock {{Growing scale-free small-world networks with tunable assortative coefficient}},
\newblock {\em Physica A}, 371, 814-822, (2006).






\end{thebibliography}
\end{document}